\documentclass[12pt]{article}
\usepackage{palatino,cite,epsfig,amsmath,amssymb}

\allowdisplaybreaks

\begin{document}

\vspace{1cm}

\begin{center}
\begin{flushright}
LAPP-EXP-2004-03\\
CERN-TH/2003-313\\
LAL 04-19
\end{flushright}
\vspace*{0.7cm}
{\Large\bf SFITTER: SUSY Parameter Analysis at LHC and LC}

\vspace{0.9cm}

{\sc
Remi Lafaye$^{1}$, Tilman Plehn$^{2}$ and Dirk Zerwas$^{3}$
}

\vspace*{0.6cm}

{\sl
$^1$LAPP-Annecy, France

\vspace*{0.3cm}

$^2$CERN Theory Division, Geneva, Switzerland

\vspace*{0.3cm}

$^3$LAL-Orsay, France
}

\end{center}

\vspace*{0.3cm}

\begin{abstract}
SFITTER is a new analysis tool to determine supersymmetric model parameters
from collider measurements. Using the set of supersymmetric mass
measurements at the LC and at the LHC we show how both colliders
probe different sectors of the MSSM Lagrangian. This observation
is a strong motivation to move from a parameter fit assuming a
certain model to the unconstrained weak-scale MSSM Lagrangian. We
argue how the technical challenges can be dealt with in a combined
fit/grid approach with full correlations.
\end{abstract}

\section{Introduction}

While the Standard Model describes all
available high energy physics experiments, it still has to be regarded
as an effective theory, valid at the weak scale. New physics are
expected to appear at the TeV energy scale. The supersymmetric
extension~(\cite{Wess:1974tw}) of the Standard Model is a well motivated extension providing us with
a description of
physics that can be extended consistently up to the unification scale.

If supersymmetry or any other high-scale extension of the Standard
Model is discovered, it will be crucial to determine its
fundamental high-scale parameters from weak-scale
measurements~\cite{Blair:2002pg}. The LHC and future Linear
Colliders will provide us with a wealth of measurements~\cite{ATLAS,Choi:2000ta}, which
due to their complexity require proper treatment to unravel the
corresponding high-scale physics. Even in the general weak-scale
minimal supersymmetric extension of the standard model (MSSM~\cite{Fayet:1974jb}) 
without any unification or SUSY breaking assumptions some of
the measurements of masses and couplings are not 
independent measurements; moreover, linking supersymmetric
particle masses to weak-scale SUSY parameters involves non-trivial
mixing to mass eigenstates in essentially every sector of the
theory. On top of that, for example in gravity mediated SUSY
breaking scenarios (mSUGRA/cMSSM) a given weak-scale SUSY
parameter will always be sensitive to several high-scale
parameters which contribute through renormalization group running.
Therefore, a fit of the model parameters using all experimental
information available will lead to the best sensitivity and make
the most efficient use of the information available.\smallskip

\begin{table}
\begin{small} \begin{center}
\begin{tabular}{|l|cccc||l|cccc|}
\hline
 & $m_{\rm SPS1a}$ & LHC & LC & LHC+LC &
 & $m_{\rm SPS1a}$ & LHC & LC & LHC+LC\\
\hline
\hline
$h$  & 111.6 & 0.25 & 0.05 & 0.05 &
$H$  & 399.6 &      & 1.5  & 1.5  \\
$A$  & 399.1 &      & 1.5  & 1.5  &
$H+$ & 407.1 &      & 1.5  & 1.5  \\
\hline
$\chi_1^0$ & 97.03 & 4.8 & 0.05  & 0.05 &
$\chi_2^0$ & 182.9 & 4.7 & 1.2   & 0.08 \\ 
$\chi_3^0$ & 349.2 &     & 4.0   & 4.0  &
$\chi_4^0$ & 370.3 & 5.1 & 4.0   & 2.3 \\
$\chi^\pm_1$ & 182.3  & & 0.55 & 0.55 &
$\chi^\pm_2$ & 370.6  & & 3.0  & 3.0 \\
\hline
$\tilde{g}$ &  615.7 & 8.0 &  & 6.5 & & & & & \\
\hline
$\tilde{t}_1$ & 411.8 &     &  2.0  & 2.0 & & & & & \\
$\tilde{b}_1$ & 520.8 & 7.5 &       & 5.7 &
$\tilde{b}_2$ & 550.4 & 7.9 &       & 6.2 \\
\hline
$\tilde{u}_1$ &  551.0 & 19.0 & & 16.0 &
$\tilde{u}_2$ &  570.8 & 17.4 & &  9.8 \\
$\tilde{d}_1$ &  549.9 & 19.0 & & 16.0 &
$\tilde{d}_2$ &  576.4 & 17.4 & &  9.8 \\
$\tilde{s}_1$ &  549.9 & 19.0 & & 16.0 &
$\tilde{s}_2$ &  576.4 & 17.4 & &  9.8 \\
$\tilde{c}_1$ &  551.0 & 19.0 & & 16.0 &
$\tilde{c}_2$ &  570.8 & 17.4 & &  9.8 \\
\hline
$\tilde{e}_1$    & 144.9    & 4.8 & 0.05 & 0.05 &
$\tilde{e}_2$    & 204.2    & 5.0 & 0.2  & 0.2  \\
$\tilde{\mu}_1$  & 144.9    & 4.8 & 0.2  & 0.2  &
$\tilde{\mu}_2$  & 204.2    & 5.0 & 0.5  & 0.5  \\
$\tilde{\tau}_1$ & 135.5    & 6.5 & 0.3  & 0.3  &
$\tilde{\tau}_2$ & 207.9    &     & 1.1  & 1.1  \\
$\tilde{\nu}_e$  & 188.2    &     & 1.2  & 1.2  & & & & & \\
\hline
\end{tabular}
\end{center} \end{small} \vspace*{-3mm}
\caption{Errors for the mass determination in SPS1a, taken from~\cite{masses:polesello}. 
Shown are the
nominal parameter values and the error for the LHC alone, the LC
alone, and a combined LHC+LC analysis. All values are given in GeV.}
\label{tab:mass-errors}
\end{table}

In a fit, the allowed parameter space might not be sampled
completely. To avoid boundaries imposed by non-physical parameter
points, which can confine the fit to a `wrong' parameter region,
combining the fit with an initial evaluation of a
multi-dimens\-ion\-al grid is the optimal approach.

In the general MSSM the weak-scale parameters can vastly
outnumber the collider measurements, so that a complete parameter fit
is not possible and one has to limit oneself to a consistent subset of
parameters. In SFITTER both grid and fit are realised and can be
combined.  This way, one can ultimately eliminate all dependence on the
starting point of the parameter determination.
SFITTER also includes a general correlation matrix and the
option to exclude parameters of the model from the fit/grid by fixing
them to a value. Additionally, SFITTER
includes the option to apply a Gaussian smearing to all observables
before they enter the fit/grid in order to simulate realistically
experimental measurements.
In this preliminary study, however, correlations and systematic
uncertainties are neglected and the central values are used for the
measurements.

Currently, SFITTER uses the predictions for the supersymmetric masses
provided by SUSPECT~\cite{Djouadi:2002ze}, but the conventions of the
SUSY Les Houches accord~\cite{Skands:2003cj} allow us to interface
other programs.  The branching ratios and $e^+e^-$ production cross
sections are provided by MSMlib~\cite{msmlib}, which has been used
extensively at LEP and cross checked with
Ref.~\cite{Barger:2001nu}. The next-to-leading order hadron collider
cross sections are computed using
PROSPINO~\cite{Beenakker:1996ch}. The fitting program uses the MINUIT
package~\cite{James:1975dr}. The determination of $\chi^2$ includes a
general correlation matrix between measurements. In its next version
SFITTER will be interfaced with the improved branching fraction
determination of SDECAY~\cite{Muhlleitner:2003vg}, as well as
alternative renormalization group codes like
SoftSUSY~\cite{Allanach:2001kg}, ISAJET~\cite{Baer:2003mg} or
SPHENO~\cite{Porod:2003um}.

\section{mSUGRA/cMSSM Parameter Determination}

\begin{table}[t]
\begin{center} \begin{small}
\begin{tabular}{|l|rr|rr|rr|rr|}
\hline
            & SPS1a & StartFit & LHC & $\Delta_{\rm LHC}$ & LC & $\Delta_{\rm LC}$ & LHC+LC & $\Delta_{\rm LHC+LC}$ \\
\hline
$m_0$       & 100 & 500 & 100.03 & 4.0  & 100.03 & 0.09 & 100.04 & 0.08 \\
$m_{1/2}$   & 250 & 500 & 249.95 & 1.8  & 250.02 & 0.13 & 250.01 & 0.11 \\
$\tan\beta$ &  10 &  50 &   9.87 & 1.3  &   9.98 & 0.14 &   9.98 & 0.14 \\
$A_0$       &-100 &   0 & -99.29 & 31.8 & -98.26 & 4.43 & -98.25 & 4.13 \\
\hline
\end{tabular}
\end{small} \end{center} \vspace*{-3mm}
\caption{Summary of the mSUGRA fits in SPS1a: true values, starting
values, fit values and absolute errors from the fit. As in SPS1a we
fix $\mu>0$. The mass values of the fits are based on can be found in
Tab.~\ref{tab:mass-errors}.}
\label{tab:msugra_fit}
\end{table}

Assuming that SUSY breaking is mediated by gravitational
interactions (mSUGRA/cMSSM) we fit four universal high-scale
parameters to a toy set of collider measurements: the universal
scalar and gaugino masses, $m_0$, $m_{1/2}$, the trilinear
coupling $A_0$ and the ratio of the Higgs vacuum expectation
values, $\tan\beta$. The sign of the Higgsino mass parameter $\mu$
is a discrete parameter and therefore fixed. In contrast to an
earlier study~\cite{Lafaye:2003} we assume the set of mass
measurement at the LHC and at the LC, shown in
Tab.~\ref{tab:mass-errors}.  The central value for our assumed
data set corresponds to the SUSY parameter point
SPS1a~\cite{Allanach:2002nj}, as computed by SUSPECT. As mentioned
in the introduction correlations, systematic errors and
theoretical errors are neglected. As the central (true) values are
used as measurements in order to study the errors on the
determination of the parameters, the $\chi^2$ values are not
meaningful and therefore are not quoted.

The starting points for the mSUGRA parameters are fixed to the mean of
the lower and upper limit (typically 1~TeV/c$^2$) of the allowed parameter range,
{\sl i.e.} they are not
necessarily close to the true SPS1a values.  The result of the fit is
shown in Tab.~\ref{tab:msugra_fit}. All true parameter values are
reconstructed well within the quoted errors, in spite of starting
values relatively far away. The measurements of $m_0$ and $m_{1/2}$
are very precise, while the sensitivity of the masses on $\tan\beta$
and $A_0$ is significantly weaker. The results for the LHC alone are
generally an order of magnitude less precise than those for the LC,
and this qualitative difference is expected to become even more pronounced
once we properly include systematical errors.

Because the data set is fit assuming mSUGRA as a unification
scenario the absence of measurements of most of the strongly
interacting particles, in particular the gluino, does not have a
strong impact on the precision of the LC determination. Therefore
the results for the combined measurements LHC+LC show only a small
improvement.

Assuming an uncorrelated data set, the correlations between the
different high-scale SUSY parameters which we obtain from the fit are
given in Tab.~\ref{tab:msugra_corr}. We can understand the correlation
matrix step by step~\cite{Drees:1995hj}: first, the universal gaugino
mass $m_{1/2}$ can be extracted very precisely from the physical
gaugino masses. The determination of the universal scalar mass $m_0$
is dominated by the weak-scale scalar particle spectrum, but in
particular the squark masses are also strongly dependent on the
universal gaugino mass, because of mixing effects in the
renormalization group running. Hence, a strong correlation between the
$m_0$ and $m_{1/2}$ occurs. The universal trilinear coupling $A_0$ can
be measured through the third generation weak-scale mass parameters
$A_{b,t,\tau}$. However, the $A_{b,t,\tau}$ which appear for example
in the off-diagonal elements of the scalar mass matrices, also depend
on $m_0$ and $m_{1/2}$, so that $A_0$ is strongly correlated with
$m_0$ and $m_{1/2}$. At this point one should stress that the
determination of $A_0$ is likely to be dominated by $A_t$ as it
appears in the calculation of the lightest Higgs mass $m_h$. After
taking into account the current theoretical error of 3~GeV on
$m_h$~\cite{schweinlein} we expect the determination of $A_0$ to
suffer significantly. The experimental errors therefore 
can be considered a call for an improvement of the 
theoretical error.\medskip

\begin{table}[t]
\begin{center} \begin{small}
\begin{tabular}{|c|rrrr|}
\hline
             &  $m_0$  &  $m_{1/2}$ & $\tan\beta$ &   $A_0$    \\
\hline
 $m_0$       & 1.000 &   -0.555 &  0.160    &  -0.324  \\
 $m_{1/2}$   &       &    1.000 &  -0.219   &   0.617  \\
 $\tan\beta$ &       &          &  1.000    &   0.307  \\
 $A_0$       &       &          &           &   1.000  \\
\hline
\end{tabular}
\end{small} \end{center} \vspace*{-3mm}
\caption[]{The (symmetric) correlation matrix for the mSUGRA fit
given in Tab.\ref{tab:msugra_fit} with data set LHC+LC.}
\label{tab:msugra_corr}
\end{table}

In general, $\tan\beta$ can be determined in three sectors of the
supersymmetric
spectrum: all four Higgs masses, and for large values of $m_A$ in
particular the light CP even Higgs mass $m_h$ depend on $\tan\beta$.
The mixing between gauginos and Higgsinos in the neutralino/chargino
sector is governed by $\tan\beta$.  Finally, the stop mixing is
governed by $\mu/\tan\beta$, while the sbottom and stau mixing depends
on $\mu \tan\beta$. The correlation of $\tan\beta$ with the other
model parameters reflects the relative impact of these three
sectors. In an earlier analysis we assumed a uniform error of 0.5\% on
all mass measurements~\cite{Lafaye:2003} and saw that in this case
$\tan\beta$ is determined through stau mixing, which in turn means
that it shows very little correlation with $m_{1/2}$.

For the more realistic scenario in Tab.~\ref{tab:mass-errors} the
outcome is the following: the relative errors for the light Higgs mass
and for the light neutralino masses at the LC are tiny. The relevant
parameter in the Higgs sector is the light stop mass, which is
governed by $m_{1/2}$; similarly the gaugino mass $m_{1/2}$ which
fixes the light neutralino and chargino masses does not depend strongly
on $\tan\beta$. The slepton sector introduces a strong correlation between
$m_0$ and $m_{1/2}$. The resulting correlation matrix is shown in
in Tab.~\ref{tab:msugra_corr}. The
results obtained with SFITTER are in agreement with expectation.

\section{General MSSM Parameter Determination}

\begin{table}
\begin{center} \begin{small}
\begin{tabular}{|l|rrr||l|rrr|}
\hline
   & AfterGrid  & AfterFit   & SPS1a & & AfterGrid  & AfterFit   & SPS1a \\
\hline
$\tan\beta$          &       100  &      10.02$\pm$3.4     &       10 &
 $M_{\tilde{u}_R}$   &      532.1 &     532.1$\pm$2.8      &    532.1 \\
$M_1$                &       100  &     102.2$\pm$0.74     &    102.2 &
 $M_{\tilde{d}_R}$   &      529.3 &     529.3$\pm$2.8      &    529.3 \\
$M_2$                &       200  &     191.79$\pm$1.9     &    191.8 &
 $M_{\tilde{c}_R}$   &      532.1 &     532.1$\pm$2.8      &    532.1 \\
$M_3$                &      589.4 &     589.4$\pm$7.0      &    589.4 &
 $M_{\tilde{s}_R}$   &      529.3 &     529.3$\pm$2.8      &    529.3 \\
$\mu$                &       300  &     344.3$\pm$1.3      &    344.3 &
 $M_{\tilde{t}_R}$   &     420.2  &     420.08$\pm$13.3    &    420.2 \\
$m_A$                &     399.35 &     399.1$\pm$1.2      &    399.1 &
 $M_{\tilde{b}_R}$   &     525.6  &     525.5$\pm$10.1     &    525.6 \\
$M_{\tilde{e}_R}$    &     138.2  &     138.2$\pm$0.76     &    138.2 &
 $M_{\tilde{q}1_L}$  &     553.7  &     553.7$\pm$2.1      &    553.7 \\
$M_{\tilde{\mu}_R}$  &     138.2  &     138.2$\pm$0.76     &    138.2 &
 $M_{\tilde{q}2_L}$  &     553.7  &     553.7$\pm$2.1      &    553.7 \\
$M_{\tilde{\tau}_R}$ &     135.5  &     135.48$\pm$2.3     &    135.5 &
 $M_{\tilde{q}3_L}$  &     501.3  &     501.42$\pm$10.     &    501.3 \\
$M_{\tilde{e}_L}$    &     198.7  &     198.7$\pm$0.68     &    198.7 &
 $A_\tau$            &    -253.5  &    -244.7$\pm$1428     &   -253.5 \\
$M_{\tilde{\mu}_L}$  &     198.7  &     198.7$\pm$0.68     &    198.7 &
 $A_t$               &    -504.9  &    -504.62$\pm$27.     &   -504.9 \\
$M_{\tilde{\tau}_L}$ &     197.8  &     197.81$\pm$0.92    &    197.8 &
 $A_b$               &    -797.99 &    -825.2$\pm$2494     &   -799.4 \\
\hline
\end{tabular}
\end{small} \end{center} \vspace*{-3mm}
\caption{Result for the general MSSM parameter determination in SPS1a
using the toy sample of all MSSM particle masses with a universal
error of 0.5\%. Shown are the nominal parameter values, the result
after the grid and the final result. All masses are given in GeV.}
\label{tab:mssm1}
\end{table}

In this study, the unconstrained weak-scale MSSM is described by
24 parameters in addition to the standard model parameters. 
The parameters are listed in Tab.~\ref{tab:mssm1}: $\tan\beta$ as in
mSUGRA, plus three soft SUSY breaking gaugino masses $M_i$, the
Higgsino mass parameter $\mu$, the pseudoscalar Higgs mass $m_A$, the
soft SUSY breaking masses for the right sfermions, $M_{\tilde{f}_R}$,
the corresponding masses for the left doublet sfermions,
$M_{\tilde{f}_L}$ and finally the trilinear couplings of the third
generation sfermions $A_{t,b,\tau}$.

\subsection{Toy model with all masses}

For testing purposes, we first consider a toy data set which includes
all supersymmetric particle masses. The universal error on all mass
measurements is set to 0.5\%.

In any MSSM spectrum, in first approximation, the parameters
$M_1$, $M_2$, $\mu$ and $\tan\beta$ determine the neutralino and
chargino masses and couplings. We exploit this feature to
illustrate the option to use a grid before starting the fit. The
starting values of the parameters other than $M_1$, $M_2$, $\mu$
and $\tan\beta$ are set to their nominal values, this study is
thus less general than the one of mSUGRA. The $\chi^2$ is then
minimized on a grid using the six chargino and neutralino masses
as measurements to determine the four parameters $M_1$, $M_2$,
$\mu$ and $\tan\beta$. The step size of the grid is 10 for
$\tan\beta$ and 100~GeV for the mass parameters.  After the
minimization, the four parameters obtained from grid minimization
are fixed and all remaining parameters are fitted. In a final run
all model parameters are released and fitted. The results after
the grid (including the complementary fit), after the final fit 
and the nominal values are shown in
Tab.~\ref{tab:mssm1}. The smearing option has not been applied.
However, the errors on the fitted values (once the fit converges)
should not be sensitive to these shortcomings.

The final fit values indeed converges to the
correct central values within its error. The central values of the fit
are in good agreement with generated values, except for the trilinear
coupling $A_{b,\tau}$. The problem is using only mass
measurements to determine the three entries in a (symmetric) scalar
mass matrix: in the light slepton sector there are three masses, left and
right scalars plus the sneutrino, so the system is in principle calculable.
In the third generation squark sector we have three
independent diagonal entries per generation and two off-diagonal
entries. But the number of mass measurements is only four, therefore
the system is underdetermined in first order. The off-diagonal entry in the mass
matrix for down type scalars includes a term $A_{b,\tau}$ and
an additional term $\mu\tan\beta$. Even for very moderate values of
$\tan\beta$ the extraction of $A_{b,\tau}$ requires precise knowledge
of $\tan\beta$. The use of branching ratios and cross section
measurements (with polarised beams) which carry information about the scalar mixing angles
should significantly improve the determination of $A_{t,b,\tau}$.

\subsection{Toy model with LHC+LC mass measurements}

\begin{table}[htb]
\begin{center} \begin{small}
\begin{tabular}{|l|rrrr|}
\hline
       & LHC & LC & LHC+LC    & SPS1a \\
\hline
$\tan\beta$          &      10.22$\pm$9.1   &    10.26$\pm$0.3  &   10.06$\pm$0.2   &       10 \\
$M_1$                &     102.45$\pm$5.3   &   102.32$\pm$0.1  &  102.23$\pm$0.1   &    102.2 \\
$M_2$                &      191.8$\pm$7.3   &   192.52$\pm$0.7  &  191.79$\pm$0.2   &    191.8 \\
$M_3$                &     578.67$\pm$15    &     fixed 500     &  588.05$\pm$11    &    589.4 \\
$M_{\tilde{\tau}_L}$ &       fixed 500      &   197.68$\pm$1.2  &  199.25$\pm$1.1   &    197.8 \\
$M_{\tilde{\tau}_R}$ &     129.03$\pm$6.9   &   135.66$\pm$0.3  &  133.35$\pm$0.6   &    135.5 \\
$M_{\tilde{\mu}_L}$  &      198.7$\pm$5.1   &    198.7$\pm$0.5  &   198.7$\pm$0.5   &    198.7 \\
$M_{\tilde{\mu}_R}$  &      138.2$\pm$5.0   &    138.2$\pm$0.2  &   138.2$\pm$0.2   &    138.2 \\
$M_{\tilde{e}_L}$    &      198.7$\pm$5.1   &    198.7$\pm$0.2  &   198.7$\pm$0.2   &    198.7 \\
$M_{\tilde{e}_R}$    &      138.2$\pm$5.0   &    138.2$\pm$0.05 &   138.2$\pm$0.05  &    138.2 \\
$M_{\tilde{q}3_L}$   &      498.3$\pm$110   &    497.6$\pm$4.4  &   521.9$\pm$39    &    501.3 \\
$M_{\tilde{t}_R}$    &       fixed 500      &      420$\pm$2.1  &  411.73$\pm$12    &    420.2 \\
$M_{\tilde{b}_R}$    &     522.26$\pm$113   &     fixed 500     &  504.35$\pm$61    &    525.6 \\
$M_{\tilde{q}2_L}$   &     550.72$\pm$13    &     fixed 500     &  553.31$\pm$5.5   &    553.7 \\
$M_{\tilde{c}_R}$    &     529.02$\pm$20    &     fixed 500     &  531.70$\pm$15    &    532.1 \\
$M_{\tilde{s}_R}$    &     526.21$\pm$20    &     fixed 500     &  528.90$\pm$15    &    529.3 \\
$M_{\tilde{q}1_L}$   &     550.72$\pm$13    &     fixed 500     &  553.32$\pm$6.5   &    553.7 \\
$M_{\tilde{u}_R}$    &     528.91$\pm$20    &     fixed 500     &  531.70$\pm$15    &    532.1 \\
$M_{\tilde{d}_R}$    &      526.2$\pm$20    &     fixed 500     &  528.90$\pm$15    &    529.3 \\
$A_\tau$             &       fixed 0        &   -202.4$\pm$89.5 &  352.1$\pm$171    &   -253.5 \\
$A_t$                &     -507.8$\pm$91    &  -501.95$\pm$2.7   & -505.24$\pm$3.3  &   -504.9 \\
$A_b$                &    -784.7$\pm$35603 &     fixed 0       &  -977$\pm$12467    &   -799.4 \\
$m_A$                &       fixed 500      &    399.1$\pm$0.9  &   399.1$\pm$0.8   &    399.1 \\
$\mu$                &     345.21$\pm$7.3   &   344.34$\pm$2.3  &   344.36$\pm$1.0  &    344.3 \\
\hline
\end{tabular}
\end{small} \end{center} \vspace*{-3mm}
\caption[]{Result for the general MSSM parameter determination in
SPS1a using the mass measurements given in
Tab.~\ref{tab:mass-errors}. Shown are the nominal parameter values and
the result after fits to the different data sets. All masses are given
in GeV.}
\label{tab:mssm2}
\end{table}

In the study of the three data sets LHC, LC, and LHC+LC in the MSSM, a fit
was performed for the data sets LHC and LC, whereas for LHC+LC additionally
the GRID was used for $M_1$, $M_2$, $\mu$ and $\tan\beta$ with the five chargino
and neutralino masses. The starting points were chosen to be the true values
(with the exception of the parameters used in the grid).
In order to obtain a solvable system,
for the LHC data set $m_A$, $M_{\tilde{t}_R}$, $M_{\tilde{\tau}_L}$, $A_\tau$ were
fixed. For the LC data set the first and second generation squark soft SUSY breaking
masses, the gluino mass $M_3$, $M_{\tilde{b}_R}$ and $A_b$ were fixed.
These parameters were chosen
on the basis of the measurements available in Tab.~\ref{tab:mass-errors}.
The values to which these parameters were fixed is not expected to influence the final
result of the fit. The results for the two data sets are shown in Tab.~\ref{tab:mssm2}.

Note that the general rule that the LHC is not sensitive to weakly interacting
particle masses is not entirely true: while the LHC has the
advantage of measuring the squark and gluino masses, the first and second
generation slepton mass parameters are also determined with a precision of the order of
percent. The results in Tab.~\ref{tab:mssm2}
show that the LHC alone is well capable of determining for example all
gaugino mass parameters as well as most of the scalar mass
parameters.

The situation at the LC is slightly different. Only marginal
information on the squark sector available at the LC. The
measurement of $A_t$ from the Higgs sector should be taken with a
grain of salt (theoretical error on the lightest Higgs mass).
Adding the stau mixing angle to the set of LC measurements will
improve the determination of $A_\tau$. However, the measurements
of the parameters, in particular slepton and gaugino parameters
are far more precise than at the LHC.

For the LHC+LC data set, a sufficient number of mass measurements is available,
so that no parameters need to be fixed. The superiority of the combination
of the measurements at the two colliders is obvious from this observation and
from Tab.~\ref{tab:mssm2}: The LHC contributes to reduce the error in the weak
sector ($M_2$) and the LC in the strongly interacting sector (third generation squarks).
Even more important: of 13 parameters undetermined by either the LHC or the LC,
11 are determined with good precision in the combination.
For $A_\tau$, $A_b$, we expect an improvement with the use of branching ratios and cross
section measurements.

A complete measurement of all parameters at the weak
scale is particularly important if one wants to probe unification scenarios
which link subsectors of the parameter space which are
independent at the weak scale.
An advanced tool like SFITTER can extract the information to probe
supersymmetry breaking scenarios from any set of measurements,
provided the set is sufficient to overconstrain the model parameters.

\section{Conclusions}

SFITTER is a new program to determine supersymmetric parameters from
measurements. The parameters can be extracted either using a fit, a
multi-dimensional grid, or a combination of the two. Correlations
between measurements can be specified and are taken into
account. While it is relatively easy to fit a fixed model with very few
parameters for example at a high scale to a set of collider
measurements, the determination of the complete set of weak-scale MSSM
model parameters requires this more advanced tool.
A mSUGRA inspired fit does not include the full complexity and power of
the combined LHC and LC data compared to the measurements at either
collider alone.
The results from SFITTER in the MSSM with the three data sets
show that only the combination of measurements of both the LHC and the LC
offers a complete picture of the MSSM model parameters
in a reasonably model independent framework.


\subsection*{Acknowledgements}

The authors would like to thank the organizers of the Les Houches
workshop and the convenors of the SUSY-group for the constructive
atmosphere in which SFITTER was born.



\end{document}